\documentclass[12pt,english,aps,manuscript]{article}
\usepackage{setspace}
\usepackage{geometry}
\usepackage{graphicx}
\usepackage{amsmath}
\usepackage{amssymb}
\usepackage{float}
\usepackage{graphics}
\usepackage{hyperref}
\usepackage{mathrsfs}

\begin{document}

\doublespacing

\def\te{\tilde e}
\def\ta{\tilde a}
\def\tl{\tilde l}
\def\tu{\tilde u}
\def\tc{\tilde c}
\def\ts{\tilde s}
\def\tb{\tilde b}
\def\tf{\tilde f}
\def\td{\tilde d}
\def\tQ{\tilde Q}
\def\tL{\tilde L}
\def\tH{\tilde H}
\def\tst{\tilde t}
\def\ttb{t\bar{t}}
\def\ttau{\tilde \tau}
\def\tmu{\tilde \mu}
\def\tg{\tilde g}
\def\tnu{\tilde\nu}
\def\tell{\tilde\ell}
\def\tq{\tilde q}
\def\tw{\widetilde W}
\def\tz{\widetilde Z}
\def\Ord{\mathcal{O}}
\newcommand{\beq}{\begin{equation}}
\newcommand{\eeq}{\end{equation}}
\newcommand {\aplt} {\ {\raise-.5ex\hbox{$\buildrel<\over\sim$}}\ }
\newcommand {\apgt} {\ {\raise-.5ex\hbox{$\buildrel>\over\sim$}}\ }
\newcommand{\Ka}{K\"ahler }



\begin{center}
\textbf{\Large Dark Matter density and the Higgs mass in LVS String Phenomenology}
\par\end{center}{\Large \par}

\begin{center}
\vspace{0.1cm}
\par\end{center}

\begin{center}
{\large Senarath de Alwis$^{\dagger}$ and Kevin Givens$^{\ddag}$ }
\par\end{center}{\large \par}

\begin{center}
Physics Department, University of Colorado \\
Boulder, CO 80309 USA
\par\end{center}

\begin{center}
\vspace{0.05cm}
\par\end{center}

\begin{center}
\textbf{Abstract}
\par\end{center}

The Large Volume Scenario for getting a non-supersymmetric vacuum in type IIB string theory leads, through the Weyl anomaly and renormalization group running, to an interesting phenomenology. However, for gravitino masses below 500 TeV there are cosmological problems and the resulting Higgs mass is well below $124$ GeV. Here we discuss the phenomenology and cosmology for gravitino masses which are $\gtrsim 500$ TeV.  We find (under some plausible cosmological assumptions) that not only is the cosmological modulus problem alleviated and the right value for dark matter density  obtained, but also the Higgs mass is in the $122$-$125$ GeV range. However the spectrum of SUSY particles will be too heavy to be observed at the LHC.
\vfill

$^{\dagger}$ {\small e-mail: dealwiss@colorado.edu}{\small \par}
$^{\ddag}$ {\small e-mail: kevin.givens@colorado.edu}{\small \par}

\eject

\vspace{5 mm}

\tableofcontents

\section{Introduction}
 The Large Volume Scenario (LVS)\cite{Bala:2005zx} of type IIB
string theory compactified on a Calabi-Yau orientifold (CYO) with
fluxes\footnote{For reviews see \cite{Grana:2005jc}\cite{Douglas:2006es}
} is a viable framework for discussing phenomenology in a top
down approach.  However the problem of getting
the MSSM in such constructions has not yet been solved. Nevertheless in
this class of models, the visible sector is localized in the CYO, so the
stabilization problem is essentially decoupled from the problem of finding
the MSSM.

By contrast, heterotic string theory models which are MSSM like
have been constructed, but the moduli stabilization problem with supersymmetry
breaking and a tunable cosmological constant is far from being solved.
In fact a major problem in these constructions is that the two issues
are not decoupled.  Given that in this case there is only one type
of flux, it is not at all clear that with current technology a solution
can be found. 

The case of type IIA strings is somewhere in between the above two
cases. While models close to the MSSM have been found (with intersecting
D6 branes for instance) and some progress on moduli stabilization
in certain special cases has been made (with supersymmetric minima),
a viable model whose SUSY breaking phenomenology can be determined is
far from being realized.

Within LVS, there have been several different versions which
in principle could have resulted in a meaningful phenomenology. However,
as has been argued by one of us in a recent paper\cite{deAlwis:2012vp}, some of these
appear to have either theoretical or phenomenological problems. A model which survives all constraints (modulo cosmological ones which are addressed here), is that discussed
in \cite{deAlwis:2009fn} and \cite{Baer:2010uy} and has been named inoAMSB.

In the following we will show that once all the phenomenological (i.e.
FCNC) and standard cosmological constraints 
are imposed, inoAMSB leads to a unique set of predictions. Our basic assumption is the following:
\begin{itemize}
\item The MSSM is located on D3 branes at a singularity of a CYO which is
of the ``Swiss Cheese'' type.
\end{itemize}

Given this assumption, we need to ensure the following theoretical
constraints in order to proceed with the LVS argument, which really
applies only to the compactified string theory in the four dimensional
low energy regime. In other words, we need to justify a 4D ${\cal N}=1$
supergravity (SUGRA) description. Flux compactifications necessarily
proceed via the ten dimensional low energy limit of string theory,
which in turn needs to result in a four dimensional theory. The constraints
follow from the requirement that the superderivative
expansion is valid at each stage. Thus we need:
\begin{itemize}
\item The energy scales of the theory $E\ll M_{KK}\ll M_{string}$.
\item After SUSY breaking $\sqrt{F}/M_{KK}\ll1$.
\end{itemize}
These are the principle theoretical constraints on the LVS construction.
In addition there are phenomenological and cosmological constraints
apart from the obvious ones, like the necessity for having a highly
suppressed cosmological constant (CC).  These constraints include:
\begin{itemize}
\item Flavor changing neutral currents (FCNC) must be suppressed.
\item The gravitino and the lightest modulus of the string theory compactification
must be heavy enough so as not to interfere with Big Bang Nucleosynthesis
(BBN) in standard cosmology.
\end{itemize}
In the context of these string constructions the FCNC constraint translates
into a lower bound on the internal volume ${\cal V}$ \cite{deAlwis:2009fn}.
The classical contribution to the soft terms at the UV scale (taken
close to the string scale) is highly suppressed (relative to the gravitino
mass) by a factor of the volume.  However, the gaugino gets a contribution
which is only suppressed by a factor of $\left(\frac{\alpha}{4\pi}\right)$, the perturbation
expansion parameter of the relevant gauge group. The soft terms at
the TeV scale are then generated by RG running essentially by the
mechanism of gaugino mediation\cite{Luty:2005sn} and are flavor neutral as
usual. So the FCNC constraint comes from comparing the classical off
diagonal contribution to the RG generated soft term, giving the relevant
lower bound.

For low values of the gravitino mass ($m_{3/2}\lesssim200\,\text{TeV}$) this
lower bound is $ $${\cal V}\gtrsim10^{5}$ in string units. The relevant
phenomenology is discussed in \cite{Baer:2010uy}. However in this
case, we have a cosmological modulus problem since for the lightest
modulus we have $m_{modulus}=m_{3/2}/\sqrt{{\cal V}}\ll10\,\text{TeV}$. Also
the neutralino contribution to dark matter density is
about an order of magnitude too low and the Higgs mass is well below
$120\,\text{GeV}$! In this note we look at the same class of models with very
high ($\gtrsim 500\,\text{TeV}$) gravitino mass. In this case the FCNC constraints are somewhat ameliorated and the CYO volume lower bound becomes ${\cal V}\gtrsim 10^4$. Up to an  $\mathcal{O}(1)$ factor, this puts
us around the lower bound for the light modulus mass but, importantly,
gives the right value for neutralino dark matter. It also gives a
Higgs mass in the $122$-$125$ GeV range in agreement with the recent
hints from CERN.

At the string  scale (assumed to be close to the GUT scale) the SUSY parameters take their values from the convential inoAMSB arguments.  The gauginos gain mass through the super-Weyl anomaly and take the following form:
\begin{equation}
M_i = \frac{b_ig_i^2}{16\pi^2}m_{3/2} \;\;\;\;\; b_i = (33/5,1,-3)
\label{weyl}\end{equation}
The scalars are suppressed relative to $m_{3/2}$ and are given by
\begin{equation}
m^2 = +\frac{3}{16}\frac{\hat{\xi}}{|\ln m_{3/2}|}\frac{m_{3/2}^2}{\mathcal{V}}  
\end{equation}
where  $\hat{\xi} \sim \Ord (1)$ is related to the Euler character of the CYO.  For $m_{3/2} = 500 \,\text{TeV} = 5\times10^{-13} \;(\text{M}_{\text{P}} = 1)$ and $\mathcal{V}= 10^4$ this gives  
\begin{equation}
m = \sqrt{\frac{3}{16}\frac{\hat{\xi}}{|\ln m_{3/2}|\mathcal{V}}}\;\;m_{3/2} \approx \sqrt{\frac{3}{16}\frac{1}{|\ln (5\!\times\!10^{-13})|\times 10^4}}\;\; 500 \; \textrm{TeV} \approx 407 \; \textrm{GeV} 
\end{equation}

The $A$, $B\mu/\mu$ and $\mu$ terms are still constrained by the size of the uplift necessary to raise the scalar potential to zero.  For $m_{3/2} = 500 \,\text{TeV}$, $\mathcal{V}= 10^4$ and $h_{21}\sim\mathcal{O}(100)$, these are
\begin{equation}
A \aplt \Ord\left(\frac{m_{3/2}}{\mathcal{V}}\right) \sim 50 \; \textrm{GeV} 
\end{equation}
\begin{equation}
\mu \sim B\mu/\mu \aplt \Ord\left(\frac{\sqrt{h_{21}}m_{3/2}}{\sqrt{\ln m_{3/2}\mathcal{V}}}\right) \sim 10 \; \textrm{TeV} 
\end{equation}

\begin{figure}[h!]
\begin{center}
\includegraphics[scale=0.82]{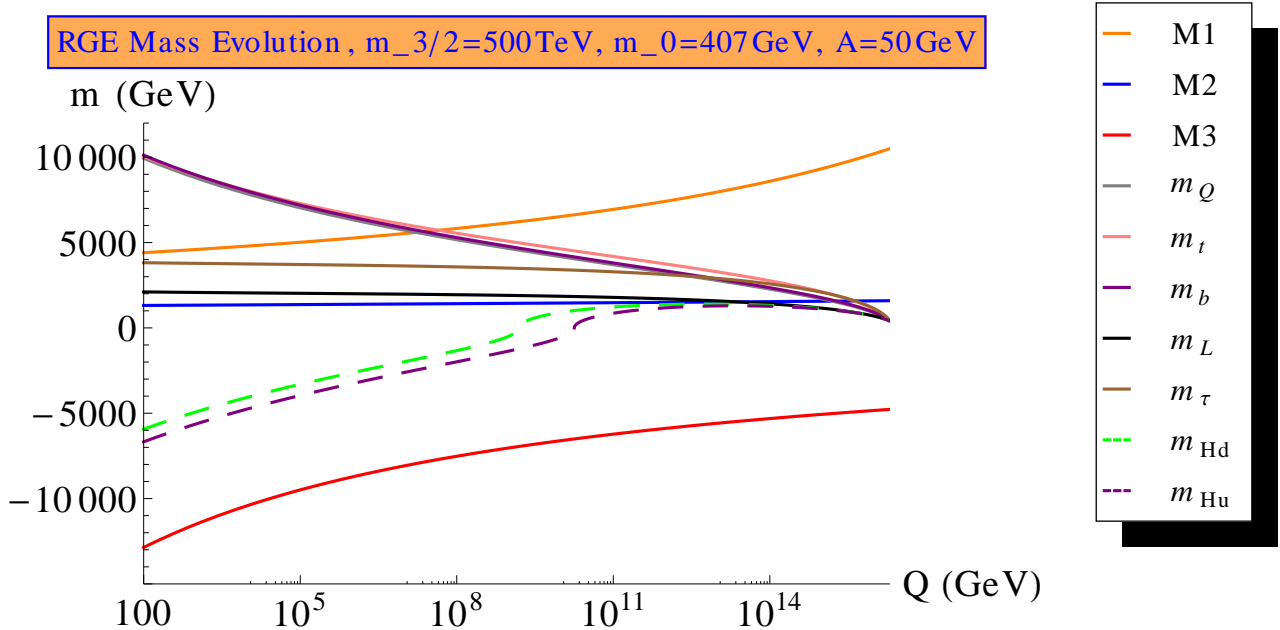}
\caption{SUSY particle mass evolution using 1-loop RGE's, $\mathcal{V} = 10^4$, $m_{3/2} = 500$ TeV, $\tan\beta = 40$, $m_0 = 407$ GeV, $A_0 = 50$ GeV}
\label{fig:RGE}
\end{center}
\end{figure}

In what follows, we compute 2-loop RGE evolution for various soft masses as well as for the $\mu$ and $B\mu/\mu$ terms using ISAJET\cite{Paige:2003mg}.  We observe that the SUSY particle masses are generically lifted to the TeV scale.  Consequently, this diminishes the likelihood of direct production at the LHC. In Table~\ref{tab:Masses}, we present the SUSY particle masses (physical mass eigenstates) for this model with various values of $m_{3/2}$ and $\tan(\beta)$.  In Figure~\ref{fig:RGE}, we plot the (1-loop) SUSY sparticle mass RGE evolution. 

In Figure~\ref{fig:RGE}, we call attention to the fact that $M^2_{H_d}$ becomes large and negative near the Weak scale.  This may seem contrary to the traditional expectation of a small and positive $M^2_{H_d}$.  However, as noted in \cite{Baer:2006rs} (page 204), for large values of $\tan(\beta)$, the bottom and tau Yukawa couplings make large contributions to the $M^2_{H_d}$ RGE, driving it negative.  For $\tan(\beta) \approx 10$,  this effect vanishes.

\begin{table}
\begin{center}
\begin{tabular}{lccccc}
\hline
parameter & inoAMSB1 & inoAMSB2 & inoAMSB3 & inoAMSB4 & inoAMSB5 \\
\hline
$m_0$       & 0 GeV & 407 GeV & 407 GeV & 615 GeV & 824 GeV \\
$m_{3/2}$   & $100,000$ & $500,000$ & $500,000$ & $750,000$ & $1,000,000$ \\
$A_0$       & 0 & 50 & 50 & 75 & 100 \\
$\tan\beta$ & 10 & 10 & 40 & 20 & 30 \\
$M_1$       & 956.1 & 4910.1 & 4918.4 & 7481.3 & 10087.0 \\
$M_2$       & 287.9 & 1400.0 & 1399.9 & 2103.7 & 2808.5 \\
$\mu$       & 1127.5 & 6465.4 & 6453.9 & 6500.3 & 8369.3 \\
$m_{\tg}$   & 2186.1 & 9501.9 & 9502.0 & 13907.0 & 18219.4 \\
$m_{\tu_L}$ & 1908.7 & 8137.9 & 8139.2 & 11824.6 & 15431.7 \\
$m_{\tu_R}$ & 1975.7 & 8485.5 & 8492.0 & 12366.6 & 16155.2 \\
$m_{\tst_1}$& 1691.8 & 7411.1 & 7135.4 & 10833.6 & 14004.7 \\
$m_{\tst_2}$& 1814.8 & 7633.6 & 7488.6 & 11004.0 & 14270.3 \\
$m_{\tb_1}$ & 1779.5 & 7602.0 & 7154.6 & 10973.1 & 14046.9 \\
$m_{\tb_2}$ & 1908.3 & 8134.8 & 7332.0 & 11665.5 & 14701.0 \\
$m_{\te_L}$ & 457.8 & 2202.3 & 2197.5 & 3286.8 & 4364.4 \\
$m_{\te_R}$ & 809.5 & 3875.2 & 3875.2 & 5789.6 & 7697.7 \\
$m_{\tw_2}$ & 1129.8 & 4599.8 & 4507.2 & 6550.3 & 8418.9 \\
$m_{\tw_1}$ & 299.7 & 1474.3 & 1474.1 & 2217.1 & 2958.5 \\
$m_{\tz_4}$ & 1143.2 & 4841.0 & 4846.4 & 7372.5 & 9939.8 \\ 
$m_{\tz_3}$ & 1135.8 & 4597.8 & 4508.3 & 6549.3 & 8421.5 \\ 
$m_{\tz_2}$ & 936.8 & 4594.6 & 4506.2 & 6548.5 & 8421.1 \\ 
$m_{\tz_1}$ & 299.4 & 1472.9 & 1471.9 & 2214.0 & 2954.3 \\ 
$m_A$       & 1208.9 & 5050.9 & 2100.4 & 6799.9 & 7253.9 \\
$m_h$       & 116.0 & 122.1 & 123.9 & 124.2 & 125.1 \\ 
$\Omega_{\tz_1}h^2$ & 0.007 & 0.111 & 0.110 &  0.111 & 0.080 \\
\hline
$\sigma\ [{\rm fb}]$ & $439$ & $6.7\times 10^{-2}$ & $7.1\times 10^{-2}$ & $2.6\times 10^{-3}$ & $8.6\times 10^{-5}$ \\
$\tg ,\tq\ {\rm pairs}$ & 3\% & 0\% & 0\% & 0\% & 0\%\\
${\rm EW-ino\  pairs}$ & 93\% & 93\% & 95\% & 96\% & 99\%\\
${\rm slep.}\ {\rm pairs}$ & 3\% & 0.5\% & 0.4\% & 0.09\% & 0.04\%\\
$\tst_1\bar{\tst}_1$ & 0\% & 0\% & 0\% & 0\% & 0\% \\
\hline
\end{tabular}
\caption{Masses and parameters in~GeV units for five case study points inoAMSB1,2,3,4,5 using ISAJET 7.79 with $m_t=172.6$ GeV and $\mu >0$. We also list the total tree level sparticle production cross section in fb at the LHC.}
\label{tab:Masses}
\end{center}
\end{table}

\section{Cosmological Issues}
\subsection*{Cosmological Modulus Problem}

Within inoAMSB models, there exists the potential for conflict with the Cosmological Modulus Problem
\cite{deCarlos:1993jw}\cite{Banks:1993en}.  Essentially, the sGoldstino, which is a light scalar modulus, can dominate the energy density of the universe and decay during the era of Big Bang Nucleosynthesis, disrupting the consistency of the BBN model.  In LVS, the light scalar modulus has a mass given by
\begin{equation}
m_{mod} \sim \frac{m_{3/2}}{\sqrt{\mathcal{V}}}
\end{equation}

For $m_{3/2} = 500 \,\text{TeV}$ and $\mathcal{V}= 10^4$, this gives a modulus mass of $m_{mod}\sim 5 \,\text{TeV}$, somewhat below the phenomenological bound of $10 \,\text{TeV}$ which is usually quoted in the literature.  We can satisfy this bound by raising $m_{3/2}$ to $1000 \,\text{TeV}$.  This will lift the sparticle spectrum into the $10's$ of TeV's, as can be seen in the last column of Table~\ref{tab:Masses}.  This greatly suppresses the likelihood of direct detection at the LHC.  In addition, one might naively assume that the dark matter relic abundance, $\Omega_{DM} h^2$, would violate known bounds.  However, as we explain later,  this bound is still satisfied. 

Nevertheless, in the case of a concrete string theory set up (e.g. LVS in type IIB), the cosmological modulus bound is actually much higher than $10 \,\text{TeV}$.  Consequently, increasing the gravitino mass to $1000 \,\text{TeV}$ does not solve the problem. The reason for this is that the standard result is derived by assuming that the coupling of the modulus to the visible sector is only suppressed gravitationally, giving a decay width of the form $\Gamma = \lambda m_{mod}^3/M_P^2$, where $\lambda \sim \Ord (1)$. However in the relevant string theory calculation there is an additional suppression  (see \cite{Cicoli:2012aq}). 
Thus the actual lower bound on the modulus mass ranges from about $30 \,\text{TeV}$ to about $100 \,\text{TeV}$ depending on the details of the scenario. The upshot is that this string theoretic picture is incompatible with the standard cosmological scenario even with a gravitino mass of $10^3 \,\text{TeV}$.  We therefore consider two possible alternatives: 

a) Thermal inflation: (see Lyth and Stewart\cite{Lyth:1995ka}). A period of late inflation (leading to just a few e-foldings of exponential expansion) appears to be very natural in the context of string theory models.  In fact as pointed out in \cite{Choi:2012ye}, this can solve the modulus problem in LVS type models even for relatively low moduli masses. On the other hand this mechanism may also dilute the dark matter abundance discussed in the next section. Thus whether or not this mechanism works in our context is dependent on the details of the model of thermal inflation. 

b) A more appealing solution within the context of the LVS scenario, in our opinion, is the one suggested by Linde\cite{Linde:1996cx}. This argument goes as follows. In general one expects in the effective action higher dimension operators of the form $\delta L = \frac{C^2}{M_P^2}s^2u^2$ where $s$ is some field which dominates the energy of the universe and $u$ is the canonically normalized (fluctuation of) the light modulus. Using the Friedman equation this gives an effective squared mass  $\Delta m^2_u=3C^2H^2$ to the field. Linde has argued that if $C\gg 1$,the amplitude of oscillations is exponentially damped by a factor $\sim \exp (-C\pi p/2)$, where $p$ is an $\Ord (1)$ number. 

Now, in the string theory context, the effective cutoff is the Kaluza-Klein scale $M_{KK}$ so that in alternative b) above $C=M_P/M_{KK}=\mathcal{V}^{2/3}$. In our LVS scenario therefore we have (for internal volume $\mathcal{V}= 10^4$) $C= 10^{8/3}$ which gives a extremely large suppression of the oscillation amplitude. This is thus a promising avenue for solving the cosmological modulus problem in our context and we hope to investigate this further in a separate publication.

\subsection*{Dark Matter Relic Abundance}

It is a well known problem\cite{Baer:2010kd} that AMSB type models generically produce a dark matter relic abundance that is several orders of magnitude lower than the current experimental value ($\Omega_{DM} h^2 \sim 0.11$\cite{Komatsu:2010fb}).  This is attributable to the near degeneracy between the lightest wino and the lightest zino.  Their mass difference is typically $\sim \Ord (100 \, \text{MeV})$.  This leads to an overabundance of $\tw$ at freeze-out and efficient $\tw$, $\tz$ co-annihilation. However, when $m_{3/2} = 500 \,\text{TeV}$, the mass difference between $\tw$ and $\tz$ is $\sim \Ord (2 \; \text{GeV})$.  This suppresses the abundance of $\tw$ at freeze-out and hence raises the SUSY contribution to dark matter relic abundance.  For $m_{3/2} = 500$ TeV, the value calculated from ISAJET is $\Omega_{DM} h^2 \approx 0.11$. 

\begin{figure}[h!]
\begin{center}
\includegraphics[scale=0.7]{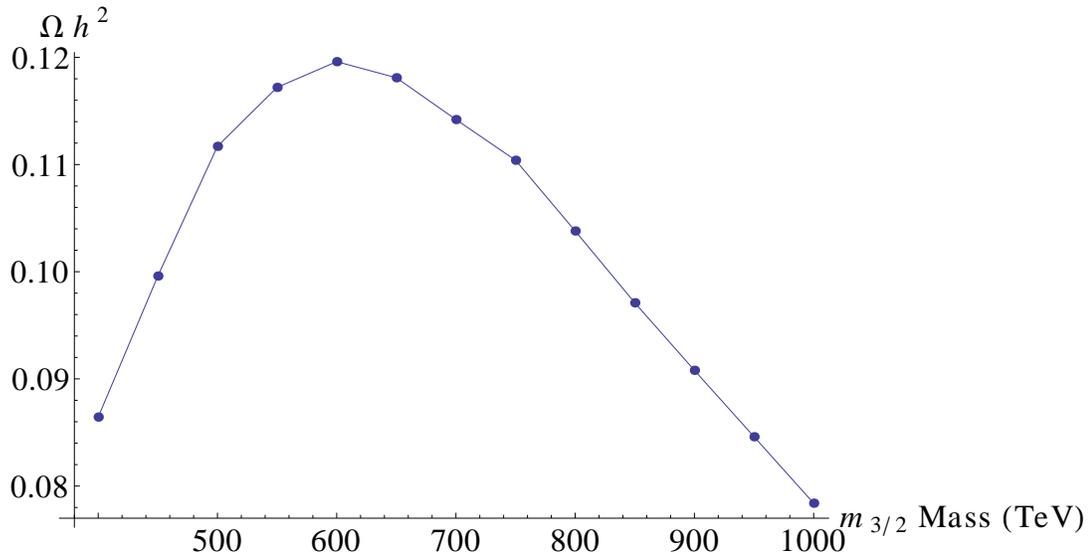}
\caption{Dark matter relic abundance as a function of $m_{3/2}$}
\label{fig:DM}
\end{center}
\end{figure}

In Figure~\ref{fig:DM}, we plot the dark matter relic abundance with $\tan(\beta) = 10$.  We see that the experimental bound is roughly saturated for $m_{3/2}= 500 \,\text{TeV}$ and  $m_{3/2}= 750 \,\text{TeV}$.  In particular, for $m_{3/2} \sim 1000 \,\text{TeV}$, the experimental bound on dark matter relic abundance is not violated but some other mechanism must account for it.  

Finally we note that in our scenario we are assuming that other possible sources of dark matter such as axions do not contribute significantly to the relic abundance.

Note that although the coannihilation process gets suppressed with increasing $m_{3/2}$, since the mass difference between the chargino and the LSP increases (see Table~\ref{tab:Masses}), the annihilation process is enhanced due to s-channel resonance effects. (For a discussion see \cite{Baer:2002fv} 
and the references to earlier work cited there). Hence, contrary to the naive expectation, at some point the dark matter density curve turns down as the gravitino mass increases beyond about $600 \,\text{GeV}$,  as seen in the figure above. 

\section{Phenomenological Issues}
\subsection*{FCNC and Anomalous Magnetic Moment of the Muon}

The SUSY particle spectrum determined by this model is also subject to constraints on known Standard Model processes.  In particular, the flavor changing neutral current process $b\!\rightarrow \!s \gamma$ as well as the anomalous magnetic moment of the muon, $\delta a_{\mu} \equiv (g-2)_{\mu}$, are influenced by the presence of SUSY particles in their respective loop diagrams.  For our model, the calculated values from ISAJET for these quantities are presented in Table~\ref{tab:Pheno} along with the corresponding experimental values\cite{Barate:1998vz}\cite{Brown:2000sj}.  From these results we conclude that our model does not violate these phenomenological constraints.     

\begin{table}[h!]
 \begin{center}
  \begin{tabular}{|r|l|l|}
      \hline
              & Experimental Value &  Estimated Value  \\
      \hline 
        $(g-2)_{\mu}$   &  $29.5 \pm 8.8 \! \times \!10^{-10}$ & $0.73 \!\times\! 10^{-10}$\\
        $BR(b \rightarrow s \gamma)$   & $3.11 \pm 0.8 \!\times\!10^{-4}$ & $3.16 \!\times\! 10^{-4}$\\
      \hline
  \end{tabular}
  \caption{Phenomenological constraints for inoAMSB with $m_{3/2} = 500$ TeV, $m_0 = 407$ GeV, $A_0 =50$ GeV, $\tan\beta = 40$ using ISAJET 7.79 with $m_t=172.6$ GeV and $\mu > 0$. }
\label{tab:Pheno}
 \end{center}
\end{table}

Note that in  Table~\ref{tab:Masses} the value of the $\mu$-term is an output determined by the experimentally measured value of the $Z$ mass. In these string theory constructions the value of this term is dependent of the mechanism which is responsible for lifting the LVS minimum to a positive value at the $10^{-3}\,\text{eV}$ scale. As discussed in \cite{deAlwis:2009fn} one needs to turn on F-terms in  either the dilaton or the complex structure directions in order to achieve this.\footnote{For an explicit example see the recent paper \cite{Cicoli:2012fh}.} Thus this term will be dependent in a complicated way on the fluxes and in general can be fine tuned to satisfy the $Z$ mass constraint. This fine tuning is of course the little hierarchy problem appearing as a landscape flux choice problem and corresponds to a fine tuning (with $m_{3/2} \sim 500\,\text{TeV} - 800\,\text{TeV}$) of 1 part in $3-4 \times 10^3$. Of course this is still much better than the original standard model fine tuning of 1 part in $10^{30}$!     

\section{Prospects for the LHC}

The fact that the SUSY particle masses are on the TeV scale will, roughly speaking, suppress the likelihood of their direct detection at the LHC.  However, there is one principle difference between this model and the case when $m_{3/2} < 100 \,\text{TeV}$. Namely, the near degeneracy between $\tw_{1}$ and $\tz_{1}$ is noticeably lifted, with $M(\tw_{1}) - M(\tz_{1}) \approx 2$ GeV.  This means that $\tw_{1}^+$ can decay to $\tz_1$ plus quarks.  However these quarks will not be energetic enough to produce jets that meet LHC trigger requirements.    

One can see from the production cross section calculations (given in the bottom rows of Table~\ref{tab:Masses}), that even for an integrated luminosity of $100 \,\text{fb}^{-1}$.  The LHC will produce fewer than $10$ events.  This clearly limits any hope of direct production of SUSY particles at the LHC.

\section{Conclusion}

As we discussed in the introduction, with a rather minimal set of string theory inputs, phenomenological constraints (chiefly the absence of FCNC) and cosmological constraints, we have obtained a very predictive phenomenology. The main output is the correlation between satisfying the light modulus and neutralino dark matter constraint on the one hand (which essentially limits the value of the gravitino mass to a range between $500 \,\text{TeV}$ and $800 \,\text{TeV}$) and the value of the mass of the light Higgs, giving the latter in the range where it may have been observed at the LHC. Unfortunately even the LSP in this scenario is at $1.4 \,\text{TeV}$, so that it is unlikely to be observed there. On the other hand, if sub TeV scale superparticles (for instance a light stop) is observed this version of string phenomenology will be ruled out.

 We should also comment here on fine-tuning or the so-called little hierarchy problem. What we would like to emphasize is that in the top down approach that we are taking there is very little freedom to get so-called natural SUSY with no (or very little) fine-tuning. While in a bottom up approach one has more freedom, the whole point of top down approaches is to find the restrictions that are imposed by the need to embed the theory in a UV complete framework such as string theory. In this case in the class of string theory models that we are considering at least, there is an inevitable little hierarchy problem (at the level of about 1 part in $10^3$). 
 
It is useful to deconstruct the arguments made here to understand precisely what input or set of inputs would need to be changed, and indeed if one has any room to maneuver whatsoever to get a low mass spectrum. Firstly, let us take a purely phenomenological supergravity approach. From this perspective the essential features of the scenario are a) sequestering i.e.\ the classical soft masses are highly suppressed compared to the gravitino mass b) the gaugino masses are generated at some high scale close to the GUT scale by the Weyl anomaly\cite{Kaplunovsky:1994fg}, i.e.\ eqn.~\eqref{weyl}. Since the actual value of the classical mass in Table~\ref{tab:Masses} is essentially irrelevant as long as it is highly suppressed compared to the gravitino mass (i.e.\ $\ll (\frac{\alpha}{4\pi})m_{3/2}$) the rest of the phenomenology and in particular the Higgs mass and the dark matter density follows.  
 
 The question is how generic is this phenomenology. Obviously the first requirement is that we start with a sequestered situation, i.e.\ one in which the classical soft parameters are highly suppressed relative to the gravitino mass. The second requirement is the validity of the Weyl anomaly formulae (eqn.~\eqref{weyl}). This follows quite generally from the Kaplunovsky-Louis formula for the gaugino mass\cite{Kaplunovsky:1994fg}\cite{Kaplunovsky:1993rd} when, as in a sequestered model, the classical term can be ignored. Then  the anomaly term $(b_i/16\pi^2)F^iK_i$ \footnote{ $K_i,~F^i$ are the derivative of the \Ka potential and the supersymmetry breaking F-term respectively.} gives  the formula eqn.~\eqref{weyl} since $F^iK_i\sim m_{3/2}$ once the CC is tuned to leading order. The string theory input, i.e.\ type IIB with LVS compactification, is simply a concrete realization of this inoAMSB framework.  
 
 Finally there is the issue of the cosmological moduli problem. As we've discussed it is plausible that it is solved either by a period of thermal inflation or by the mechanism discussed in \cite{Linde:1996cx}. Thus our conclusions are based on the following two assumptions about the cosmology of our model.

 1. Dark matter consists  mainly of thermal LSP's and other possible candidates such as axions do not give significant contributions to the relic density even in the absence of a dilution mechanism.
 
 2. The cosmological moduli problem can be solved either by a mechanism similar to that suggested in \cite{Linde:1996cx} or (less plausibly) by a period of thermal inflation which while diluting the light modulus does not seriously affect the dark matter abundance.

 If these two assumptions are satisfied then our string theoretic model gives a Higgs mass that is consistent with the observed value and a dark matter density of the right amount, when the gravitino mass is taken to be in the range $500-750 \,\text{TeV}$.

 \section{Acknowledgements} We wish to thank Howie Baer for his comments on the manuscript and his assistance in running the ISAJET program.  SdA would also like to thank Kiwoon Choi for a discussion on thermal inflation and the modulus problem in LVS and Michele Cicoli for discussions on \cite{Cicoli:2012aq} . We also wish to thank an anonymous referee for drawing are attention to \cite{Cicoli:2012aq} and for several incisive comments. This research is partially supported by the United States Department of Energy under grant DE-FG02-91-ER-40672. 
 
\bibliographystyle{hieeetr}      
\bibliography{500refs}		 
\end{document}